\documentclass[11pt,a4paper]{article}
\pdfoutput=1
\usepackage{jcappub}
\usepackage{graphicx}
\usepackage{longtable}
\usepackage{amsmath}
\usepackage{amssymb}
\usepackage{subfigure}
\usepackage{hyperref}
\usepackage{dcolumn}
\usepackage{multirow}
\usepackage{cleveref}
\usepackage{bm}
\usepackage{color}

\newcommand{\be}{\begin{eqnarray}}
\newcommand{\ee}{\end{eqnarray}}
\newcommand{\mpl}{M_{\rm {pl}}}

\newcommand{\dd}{\, {\rm d}}
\newcommand{\gsim}{\;\mbox{\raisebox{-0.5ex}{$\stackrel{>}{\scriptstyle{\sim}}$}
}\;}
\newcommand{\lsim}{\;\mbox{\raisebox{-0.5ex}{$\stackrel{<}{\scriptstyle{\sim}}$}
}\;}
\def\eea{\end{eqnarray}}
\def\bea{\begin{eqnarray}}

\newcommand{\ve}{\varepsilon}

\newcommand{\rt}{\tilde{r}}

\newcommand{\U}{\Upsilon}
 
 \newcommand{\la}{\lambda}

\newcommand{\s}{\sigma}

\newcommand{\nm}{{\mu\nu}}

\newcommand{\rpp}{r'}
\newcommand{\GN}{G_{\rm N}}

\newcommand{\vpp}{\varphi}

\newcommand{\dn}{\delta\nu}
\newcommand{\dl}{\delta\lambda}
\newcommand{\oo}{\mathcal{O}}

\newcommand{\rv}{r_{\rm v}}

\newcommand{\lag}{\mathcal{L}}
\newcommand{\uo}{\Upsilon_1}

\newcommand{\ut}{\Upsilon_2}
\newcommand{\pdot}{v_0}

\newcommand{\C}{{\cal M}}

\allowdisplaybreaks
\begin{document}
\title{Relativistic Stars in Beyond Horndeski Theories}
\author[a]{Eugeny Babichev}
\emailAdd{eugeny.babichev@th.u-psud.fr}
\affiliation[a]{Laboratoire de Physique Th\'eorique, CNRS, Univ.~Paris-Sud, Universit\'e Paris-Saclay, 91405 Orsay, France}
\author[b]{Kazuya Koyama}
\emailAdd{kazuya.koyama@port.ac.uk}
\affiliation[b]{Institute of Cosmology and Gravitation, University of Portsmouth, Portsmouth PO1 3FX, UK}
\author[c]{David Langlois}
\emailAdd{langlois@apc.univ-paris7.fr}
\affiliation[c]{Laboratoire APC -- Astroparticules et Cosmologie, Universit\'e Paris Diderot Paris 7, 75013 Paris, France}
\author[d]{Ryo Saito}
\emailAdd{rsaito@yukawa.kyoto-u.ac.jp}
\affiliation[d]{Center for Gravitational Physics, Yukawa Institute for Theoretical Physics, Kyoto University, 606-8502, Kyoto, Japan}
\author[b]{Jeremy Sakstein}
\emailAdd{jeremy.sakstein@port.ac.uk}
\abstract{This work studies relativistic stars in beyond Horndeski scalar-tensor theories that exhibit a breaking of the Vainshtein mechanism inside matter, focusing on a model based on the quartic beyond Horndeski Lagrangian. We self-consistently derive the scalar field profile for static spherically symmetric objects in asymptotically de Sitter space-time and show that the Vainshtein breaking branch of the solutions is the physical branch thereby resolving several ambiguities with non-relativistic frameworks. 
The geometry outside the star is shown to be exactly Schwarzschild-de Sitter and therefore the PPN parameter $\beta_{\rm PPN}=1$, confirming that the external screening works at the post-Newtonian level. The Tolman-Oppenheimer-Volkoff (TOV) equations are derived and a new lower bound on the Vainshtein breaking parameter $\Upsilon_1>-4/9$ is found by requiring the existence of static spherically symmetric stars. Focusing on the unconstrained case where $\Upsilon_1<0$, we numerically solve the TOV equations for polytropic and realistic equations of state and find stars with larger radii at fixed mass. Furthermore, the maximum mass can increase dramatically and stars with masses in excess of $3M_\odot$ can be found for relatively small values of the Vainshtein breaking parameter. We re-examine white dwarf stars and show that post-Newtonian corrections are important in beyond Horndeski theories and therefore the bounds coming from previous analyses should be revisited. }
\maketitle

\section{Introduction}

The poorly understood mechanism driving the acceleration of the cosmic expansion---the so-called dark energy pervading our universe---has led to a recent effort into the study of alternative theories of gravity as a possible explanation (see \cite{Clifton:2011jh,Joyce:2014kja,Koyama:2015vza,Bull:2015stt} for reviews). Many of the proposed alternatives are scalar-tensor theories which include one 
extra scalar degree of freedom that interacts with the metric and drives the cosmic expansion. 

In particular, recent developments have led to the exploration of  scalar-tensor theories involving  higher-order derivatives of a scalar field in their Lagrangians. Although these typically lead to the introduction of an Ostrogradsky ghost-like degree of freedom, it is nevertheless possible to construct classes of ghost-free Lagrangians of this type that propagate three degrees of freedom (i.e. two tensor modes and a single scalar mode). Early models, known as Horndeski models~\cite{Horndeski:1974wa} (see also \cite{Deffayet:2011gz}), were constructed by requiring the equations of motion to be manifestly second order, but it was lately realized that this condition is in fact not necessary to avoid ghosts, leading to theories ``beyond Horndeski'' \cite{Zumalacarregui:2013pma,Gleyzes:2014dya,Gleyzes:2014qga}. As understood very recently, a crucial ingredient to avoid the Ostrogradsky ghost is the degeneracy of the total Lagrangian, taking into account the coupling between scalar field and metric \cite{Langlois:2015cwa,Langlois:2015skt} (see also \cite{Motohashi:2016ftl} and \cite{Klein:2016aiq} in the context of classical mechanics). Imposing the degeneracy of the Lagrangian in a systematic way, new scalar-tensor theories were uncovered in \cite{Langlois:2015cwa}, in addition to the Horndeski and beyond Horndeski theories previously known. 
All these theories, in particular how they change under disformal transformations of the metric, were further studied in \cite{Crisostomi:2016tcp,Crisostomi:2016czh,deRham:2016wji,Achour:2016rkg}.
 
In some theories of modified gravity, such as Horndeski theories, consistency with the predictions of general relativity (GR) can be achieved on small scales thanks to the Vainshtein mechanism \cite{Vainshtein:1972sx} (see \cite{Babichev:2013usa,Khoury:2013tda} for reviews), which utilities the higher-derivative terms in the equations of motion to suppress the scalar field gradient sourced by massive objects. Indeed, expanding the metric sourced by an object of mass $M$ to Newtonian order as
\begin{equation}\label{eq:iso}
\dd s^2 =(-1+2\Phi)\dd t^2 + (1+2\Psi)\, \delta_{ij}\dd x^i\dd x^j \,, 
\end{equation}
one finds a correction to the Newtonian potential
\begin{equation}
\frac{\dd\Phi}{\dd r}=\frac{G_N M}{r}\left[1+2\alpha^2\left(\frac{r}{\rv}\right)^n\right],
\end{equation}
where the dimensionless constant $\alpha$ parameterises the coupling of the scalar to matter and $n>0$ is model dependent. A solar mass object has $\rv\sim \oo(0.1\textrm{ kpc})$ \cite{Khoury:2013tda} and so the correction to GR is strongly suppressed in the solar system. 
In the case of Horndeski theories, Vainshtein screening is 
fully effective \cite{Koyama:2013paa, Kimura:2011dc, Narikawa:2013pjr}. 
For beyond Horndeski theories, this mechanism works outside extended bodies but breaks down {inside} matter \cite{Kobayashi:2014ida}. The equations governing Newtonian perturbations were found to be of the form \cite{Kobayashi:2014ida,Koyama:2015oma,Saito:2015fza,Sakstein:2016ggl}
\begin{align}
\frac{\dd\Phi}{\dd r}&=\frac{\GN M(r)}{r^2}+\frac{\uo \GN}{4}\frac{\dd^2M(r)}{\dd r^2}\\
\frac{\dd\Psi}{\dd r}&=\frac{\GN M(r)}{r^2}-\frac{5\ut\GN}{4r}\frac{\dd M(r)}{\dd r}\,,
\end{align}
where $M(r)\equiv 4\pi\int_0^r s^{2}\rho(s)ds$, and the parameters $\uo$ and $\ut$ are non vanishing when the theory contains beyond Horndeski terms in its Lagrangian.

This opens up the possibility of testing beyond Horndeski theories using astrophysical objects such as stars \cite{Koyama:2015oma,Saito:2015fza,Sakstein:2015zoa,Sakstein:2015aac,Jain:2015edg} and galaxy clusters \cite{Sakstein:2016ggl}. Currently, $\uo$ is bounded in the range $-0.22<\uo<0.027$ where the lower bound comes from the Chandrasekhar mass of white dwarf stars \cite{Jain:2015edg} and the upper bound comes from consistency of the minimum mass for hydrogen burning with the lowest mass hydrogen burning star \cite{Sakstein:2015zoa,Sakstein:2015aac}. For later purposes we note that prior to the white dwarf constraint, Ref. \cite{Saito:2015fza} was able to place the lower limit $\uo>-2/3$ by requiring a sensible stellar profile (with a mass density that decreases with the radius). The best constraint on $\ut=-0.22^{+1.22}_{-1.19}$ comes from the agreement of the lensing and hydrostatic mass of galaxy clusters \cite{Sakstein:2016ggl}. 

Constraining these parameters is important because they are directly related to the coefficients introduced in the context of the effective description of dark energy that includes Horndeski and beyond Horndeski theories \cite{Gleyzes:2013ooa,Bellini:2014fua,Gleyzes:2014rba}, via \cite{Saito:2015fza,Sakstein:2016ggl}:
\begin{equation}\label{eq:EFT}
\Upsilon_1=\frac{4\alpha_H^2}{c_T^2(1+\alpha_B)-\alpha_H-1}\quad\textrm{and}\quad \Upsilon_2 = \frac{4\alpha_H(\alpha_H-\alpha_B)}{5(c_T^2(1+\alpha_B)-\alpha_H-1)}.
\end{equation}
The coefficients $\alpha_T\equiv c_T^2-1$, $\alpha_B$ and $\alpha_H$ are defined at the level of the cosmological background solution and characterise the behaviour of cosmological perturbations \cite{Gleyzes:2014rba}. In particular, when the theory is purely Horndeski $\alpha_H=0$ and we thus have $\uo=\ut=0$. Therefore, constraints on $\Upsilon_i$ directly restrict the allowed ``beyond Horndeski'' deviations from GR.

The constraints mentioned above all rely on non-relativistic systems. The purpose of this paper is to investigate the existence and structure of relativistic stars in these theories. There are several motivations for such a study. First, the equations of motion for beyond Horndeski theories are very non-linear and it is important to verify that static spherically symmetric solutions for relativistic stars exist. Second, there are technical issues relating to the cosmological matching of the small-scale scalar field profile. The approach taken by \cite{Kobayashi:2014ida,Koyama:2015oma,Jimenez:2015bwa} was to perturb the metric potentials and the scalar field about their cosmological values and then take the non-relativistic limit. The resulting scalar equation of motion is cubic and so one has to choose the correct branch. Unfortunately, knowing which one is correct requires knowledge of Hubble-scale corrections, which are absent in this non-relativistic treatment. There are three branches of solutions, the Vainshtein breaking one where GR is recovered outside astrophysical bodies but deviations are present inside and a second one where inverse-square potentials are obtained both inside and outside the source but deviations from GR are present because the Eddington light bending parameter $\gamma_{\rm PPN}\ne 1$. Discerning the correct branch is very important since different works in the literature use different branches to place constraints\footnote{For example \cite{Koyama:2015oma,Saito:2015fza,Sakstein:2015zoa,Sakstein:2015aac,Jain:2015edg,Sakstein:2016ggl} use the Vainshtein breaking branch whereas \cite{Jimenez:2015bwa} use the other one.}. Third, neutron stars could serve as a new and novel probe of these theories. Indeed, the study of relativistic objects in other higher-derivative theories has proved fruitful \cite{Babichev:2012re,Hui:2012jb,Hui:2012qt,Babichev:2013cya,Sotiriou:2013qea,Sotiriou:2014pfa,Chagoya:2014fza,Kobayashi:2014eva,Cisterna:2015yla,Maselli:2015yva,Appleby:2015ysa,Cisterna:2016vdx,Maselli:2016gxk,Silva:2016smx,Minamitsuji:2016hkk}. Finally, white dwarf stars have been used to place new bounds on beyond Horndeski theories using the non-relativistic equations \cite{Jain:2015edg} but the importance of post-Newtonian corrections has not yet been explored. Our analysis allows us to perform such an exploration and we find that post-Newtonian corrections are indeed important.

The importance of the asymptotic boundary conditions was previously emphasised in the context of the cubic galileon by \cite{Babichev:2012re} where it was shown that imposing Minkowski or de Sitter asymptotic conditions picks out different branches of solutions. In the present work, we follow the approach of \cite{Babichev:2012re} to keep track of Hubble-scale corrections. We consider 
a model whose Lagrangian contains the quartic beyond Horndeski term and a standard kinetic term for the scalar field, as well as cosmological constant. We study 
a static, spherically symmetric object embedded in de Sitter spacetime. By transforming from Friedmann-Lema\^{i}tre-Robertson-Walker (FLRW) coordinates to the Schwarzschild slicing of de Sitter we are able to obtain both the weak-field limit and the Tolman-Oppenheimer-Volkoff (TOV) system whilst simultaneously being able to keep track of Hubble-scale corrections and without any ambiguities coming from matching different coordinates. 

The main results of this paper are the following:
\begin{itemize}
\item We recover the weak-field limit from a fully relativistic approach and show that the branch of solutions that matches onto asymptotic de-Sitter space is the Vainshtein breaking branch.
\item We find an exact vacuum solution for the space-time exterior to the star, which allows us to derive the parameterised post-Newtonian (PPN) parameter $\beta_{\rm PPN}=1$ so that the theory behaves like GR to post-Newtonian order, at least when it comes to the precession of the orbits of celestial bodies.
\item We derive the TOV system of equations governing the structure of relativistic stars and integrate them numerically for polytropic and realistic nuclear equations of state. We find physically acceptable
configurations that are compatible with observations, although for large negative values of $\Upsilon_1$ one finds large radii compared with GR for a fixed equation of state (EOS) and maximum masses in excess of $2.5M_\odot$.
\item We find a relativistic correction to the existence condition raising the lower bound from $\uo>-2/3$ to $\uo>-4/9$.
\item We re-investigate white dwarf stars using the full TOV equations and find that post-Newtonian corrections are important for massive stars, so much so that the Chandrasekhar mass for $\U_1<0$ is larger than the GR prediction, in contrast to the non-relativistic case. For this reason, the bounds found using white dwarf stars should be revisited \cite{Jain:2015edg}.
\end{itemize}

The paper is organised as follows: we first present a model that exhibits Vainshtein breaking and study its cosmology in FLRW coordinates in \cref{sec:modcos}, focusing on exact de Sitter solutions, which allows us to perform an exact transformation to Schwarzschild-like coordinates. In \cref{sec:ss} we 
examine the structure of static spherically symmetric objects. The sub-horizon weak-field limit is reviewed in order to remind the reader of the ambiguities associated with selecting a branch. The values of $\GN$ and $\gamma_{\rm PPN}$ ($=1$) are derived and are found to agree with the non-relativistic treatment. Next, we focus on the full relativistic problem and find an exact solution for the metric exterior to the star. Using this, we show that $\beta_{\rm PPN}=1$ and that the Vainshtein breaking solution is the one which has the correct asymptotic limit i.e. that space-time is asymptotically de Sitter. Finally, we derive and numerically solve the TOV system for relativistic stars using polytropic and realistic equations of state. We discuss our results and conclude in \cref{sec:concs}.

\section{Model and cosmological de Sitter solution}\label{sec:modcos}

For simplicity and concreteness, we will study one of the simplest models which exhibits Vainshtein breaking inside matter\footnote{This model is free from the conical singularity that can appear in a special subclass of models investigated in \cite{DeFelice:2015sya,Kase:2015gxi}. Note that, in those papers,  $\alpha_H$ is defined as a local function and thus coincides only asymptotically with the standard definition of $\alpha_H$, which depends only on the homogeneous cosmological solution.}, characterised by the action 
\begin{equation}
\label{action}
S=\int d^4x\, \sqrt{-g}\, \left[
\mpl^2\left(\frac{R}{2}-\Lambda\right)-k_2\mathcal{L}_2+f_4\lag_{4,{\rm bH}}\right]\,,
\end{equation}
with
\begin{align}
\mathcal{L}_2&= \phi_{\mu}\phi^{\mu} \equiv X\,\\
\mathcal{L}_{4,{\rm bH}}&=-X\left[(\Box\phi)^2-(\phi_{\nm})^2\right]+2\phi^\mu\phi^\nu\left[\phi_\nm\Box\phi-\phi_{\mu\sigma}\phi^{\sigma}_{\,\,\nu}\right],
\end{align}
where $\Lambda$ is a (positive) cosmological constant and $k_2$ and $f_4$ are constant coefficients. 
Here, we have used the shorthand notations, $\phi_{\mu} \equiv \nabla_{\mu}\phi$ and $\phi_{\mu\nu} \equiv \nabla_{\mu}\nabla_{\nu}\phi$.
We note that $\mpl^2=(8\pi G)^{-1}$ where $G$ is not Newton's constant $\GN$ but must be related to it by matching to the weak field limit. The Lagrangian $\mathcal{L}_{4,{\rm bH}}$ is one of the two beyond Horndeski terms introduced in \cite{Gleyzes:2014dya}, which lead to higher order equations of motion but without suffering from an Ostrogradsky instability. The theory (\ref{action}) contains two tensor modes and a single scalar mode, as can be deduced from the general Hamiltonian analysis of \cite{Langlois:2015skt}\footnote{Another Hamiltonian analysis, but restricted to $\mathcal{L}_{4,{\rm bH}}$, was presented in \cite{Deffayet:2015qwa}, with the conclusion that the total number of degrees of freedom was strictly less than four.}. Note that (\ref{action}) corresponds to the model studied by \cite{Koyama:2015oma} augmented by a cosmological constant. 

Matter, characterised by the energy-momentum tensor $T_{\mu\nu}$, is assumed to be minimally coupled to the metric $g_{\mu\nu}$ that appears in the action (\ref{action}). As a consequence, the energy-momentum tensor satisfies the usual conservation equation
\begin{equation}
\nabla_\mu T^{\mu}_{\ \nu}=0\,.
\end{equation}
The tensor equations of motion, which generalise Einstein's equations, can be written in the form
\begin{equation}
\mpl^2(G_{\mu\nu}+\Lambda g_{\mu\nu})+ H_{\mu\nu}= T_{\mu\nu}\,,
\end{equation}
where $G_{\mu\nu}$ is the familiar Einstein tensor and $H_{\mu\nu}$ represents the new terms derived 
from $\mathcal{L}_2$ and $\mathcal{L}_{4,{\rm bH}}$. 
Finally, since the scalar sector of the theory is shift-symmetric, the equation of motion for the scalar field can be written in the form 
\begin{equation}\label{eomphi}
\nabla_\mu J^\mu=0\,.
\end{equation}
The explicit expressions for $H_{\mu\nu}$ and $J^\mu$ are rather involved and we will not write their general form here but simply give their relevant components in a static spherical symmetric geometry (see \cite{Kobayashi:2014ida} for the general equation in beyond Horndeski theories). 

We now seek vacuum (i.e. no matter energy-momentum tensor in addition to the cosmological constant) de Sitter cosmological solutions, expressed in FLRW coordinates, 
\begin{equation}
\label{metric_dS}
\dd s^2 = -\dd \tau^2 + e^{2H\tau} \left(\dd\rpp^2 + \rpp^2\dd\Omega_2^2 \right) \,, 
\end{equation}
with $H$ constant. The scalar equation of motion reduces to 
\begin{equation}
\partial_\tau (a^3 J^\tau)=0,
\end{equation}
which is solved by $J^\tau=0$ (the general solution $J^\tau\propto a^{-3}$ quickly approaches this particular solution). 

Substituting the explicit expression for the current, one gets the equation
\begin{equation}\label{eq:bgeomphi}
J^\tau=-2\dot{\phi}\left(k_2 + 12 f_4 H^2 \dot{\phi}^2\right)=0\,.
\end{equation}
Einstein's equations give the Friedmann constraint, which reads
\begin{equation}\label{eq:fried}
3 \mpl^2 H^2=\mpl^2\Lambda + k_2 \dot{\phi}^2 +30 f_4 H^2 \dot{\phi}^4,
\end{equation}
where an over-dot denotes a derivative with respect to cosmic time $\tau$. 

Replacing $\dot\phi$ by $v_0$ and introducing the dimensionless quantity $\s^2\equiv \Lambda/(3 \mpl^2 H^2)$, one finds that the two above equations imply 
\begin{equation}
k_2=-2\frac{\mpl^2H^2}{v_0^2}\left(1-\sigma^2\right)\,,\qquad 
f_4=\frac{\mpl^2}{6v_0^4}\left(1-\s^2\right)\,.\label{eq:f4k2rel}
\end{equation}
In what follows, we will always eliminate $k_2$ and $f_4$ in favour of $\pdot$, $H$ and $\s$. This will guarantee that our local solution is related to a well defined cosmological solution asymptotically.

The FLRW slicing of de Sitter spacetime is not well adapted to study static spherically symmetric objects such as stars. It is therefore convenient to work in Schwarzschild-like coordinates using the transformation
\begin{align}
\tau=t+\frac{1}{2H}\ln\left[1-H^2r^2\right]\quad\textrm{and}\quad\rpp=\frac{e^{-Ht}}{\sqrt{1-H^2r^2}}\, r\,.
\end{align}
In terms of the new coordinates $t$ and $r$, the metric (\ref{metric_dS}) reads
\begin{equation}
\label{metric_c}
 \dd s^2=-(1-H^2r^2)\dd t^2+\frac{\dd r^2}{1-H^2r^2}\dd r^2 +r^2\dd\Omega_2^2 \,,
\end{equation}
while the scalar field cosmological solution becomes
\begin{equation}
\label{sf_c}
 \phi(r,t)=v_0t+\frac{v_0}{2H}\ln\left(1-H^2r^2\right)\,,
\end{equation}
which now depends on both temporal and radial coordinates. One may check that these expressions solve the current and tensor equations in this coordinate system.

\section{Static Spherically Symmetric Objects}\label{sec:ss}
We now introduce an astrophysical object, which we model as a spherical symmetric perfect fluid configuration whose energy-momentum 
tensor is of the form
\begin{equation}
 T^\mu_\nu=\textrm{diag}\left(-\varepsilon,P,P,P\right)\,,
\end{equation}
where $\ve(r)$ and $P(r)$ denote the energy density and pressure, respectively. Introducing this source modifies the spacetime metric, which we now write as
\begin{equation}
 \dd s^2 = -e^{\nu(r)}\dd t^2 + e^{\lambda(r)}\dd r^2 + r^2\dd\Omega_2^2\,.
\end{equation}

The relevant equations of motion are the following: one needs the time and radial components of the tensor equations of motion, which are
\begin{eqnarray}
\label{E00}
\ve&=&\mpl^2\frac{e^{-\lambda}}{r^2}\left(-1+e^\lambda+r\lambda'\right)-\mpl^2\Lambda-k_2 \left(e^{-\nu} v_0^2+e^{-\lambda}\phi^{\prime 2}\right)
\cr
&& +f_4\left[\frac{e^{-\nu-2\lambda}}{r^2}v_0^2\left((-10+14r\lambda')\phi^{\prime 2}-20 r\phi'\phi''\right)+\frac{e^{-3\lambda}}{r^2} \left((2-10r\lambda') \phi^{\prime 4}+16 r \phi^{\prime 3}\phi''\right)
\right]\,,\qquad 
\\
\label{Err}
P &=& \mpl^2\frac{e^{-\lambda}}{r^2}\left(1-e^\lambda+r\nu'\right)+\mpl^2\Lambda
-k_2 \left(e^{-\nu} v_0^2+e^{-\lambda}\phi^{\prime 2}\right)
\cr
&& +f_4\left[\frac{e^{-\nu-2\lambda}}{r^2}v_0^2\left((2+14r\nu')\phi^{\prime 2}+4 r\phi'\phi''\right)-10\frac{e^{-3\lambda}}{r^2} \phi^{\prime 4}\left(1+r\nu'\right)
\right]\,,\qquad 
\end{eqnarray}
as well as the equation of motion for the scalar field, Eq.~(\ref{eomphi}), which reduces to 
\begin{equation}
\partial_r\left[r^2 e^{(\nu+\lambda)/2}J^r\right]=0\,,
\end{equation}
implying that $J^r=0$. Substituting the explicit expression of the radial component of the current, we get the equation
\begin{align}\label{eq:je0}
J^r&=\frac{8f_4e^{-3\la}}{r^2}\left[1+r\nu'\right]\phi'^3+2e^{-2\la-\nu}\left[k_2 e^{\la+\nu} - f_4v_0^2\frac{5\nu'+\lambda'}{r}\right]\phi'=0\,.
\end{align} 
Note that if $J^r=0$ the time-radial component of the metric equations of motion is automatically satisfied
since it is proportional to $J^r$ for the ansatz assumed in this paper~\cite{Babichev:2015rva}. Finally, the energy-momentum tensor $\nabla_\mu T^\nm=0$ yields
\begin{equation}\label{eq:emcons}
 \nu'=\frac{2P'}{\varepsilon+P},
\end{equation}
where a prime denotes a derivative with respect to $r$.

Far from the star the solution must asymptotically approach the cosmological solution (\ref{metric_c})-(\ref{sf_c}):
\begin{equation}
\label{asymp_dS}
 \nu\approx -\lambda\sim \ln\left(1-H^2r^2\right),\quad
 \phi \sim v_0t+\frac{v_0}{2H}\ln\left(1-H^2r^2\right) \qquad \textrm{for} \quad r_*\ll r < H^{-1}\,.
\end{equation}
Note that the coordinate $r$ is bounded by the value $r_{\rm H}=H^{-1}$, corresponding to de Sitter horizon.

\subsection{Sub-Horizon Weak-Field Limit}\label{ss:subweaklimit}

In order to examine the sub-horizon weak-field limit we expand the metric potentials and scalar as
\begin{align}
 \nu(r)&=\ln\left(1-H^2r^2\right)+\delta\nu(r)\,, \label{eq:nusubh} \\
 \lambda(r)&=-\ln\left(1-H^2r^2\right)+\dl(r) \label{eq:lamsubh} \quad \textrm{and} \\
 \phi(r,t) &= v_0t+\frac{v_0}{2H}\ln\left(1-H^2r^2\right)+\varphi(r)\,,\label{eq:phisubh}
\end{align}
where we expect $\dn\sim\dl\sim G_N M/r\ll1$ for an object of mass $M$. 
 Furthermore, we assume that the cosmological corrections that depend on the small (in the sub-horizon limit) parameter $Hr\ll 1$ are negligible with respect to the perturbations due to the central object, namely we assume $H^2r^2 \ll \delta \nu, \, \delta \lambda$ and $v_0Hr^2/2 \ll \varphi$ (see \cref{app:cosmoterms}). These assumptions are valid at sufficiently small radii; using the results obtained below, one can show that this is true for $r \ll (G_N M/H^2)^{1/3}$, which is much larger than the stellar radius. 
Simplifications for the scalar field are not so straightforward because non-linearities due to higher derivatives may be important and so we retain all non-linear terms that are not suppressed by powers of $\dl$ or $\dn$, and ignore terms such as $\vpp'^4/\pdot^4$ compared with $\vpp'^2/\pdot^2$ and $r\vpp'\vpp''/\pdot^2$ since they are suppressed by extra powers of $\vpp'/\pdot$ (see \cite{Kobayashi:2014ida} for a discussion on this). 

With these simplifications, the $00$-component of the tensor equation of motion, Eq.~(\ref{E00}), becomes 
\begin{equation}\label{eq:pertE00}
8\pi G r^2\varepsilon=\dl + r\dl' -\frac{5(1-\s^2)\vpp'^2}{3v_0^2}-\frac{10(1-\s^2)r\vpp'\vpp''}{3v_0^2},
\end{equation}
which can be integrated once to give
\begin{equation}
\dl=\frac{2GM}{r}+\frac{5(1-\s^2)\vpp'^2}{3v_0^2}\,,
\end{equation}
where we have introduced the function 
\begin{equation}
M(r)\equiv 4\pi\int dr\, r^2\ve\,,
\end{equation}
corresponding to the mass within the radius $r$.

Substituting the above expression for $\dl$ into the 
{simplified} $rr$-component, Eq.~(\ref{Err}), one gets 
\begin{equation}\label{eq:pertErr}
\dn'=\frac{2GM}{r^2}+\frac{4(1-\s^2)\vpp'^2}{3v_0^2r}-\frac{2(1-\s^2)\vpp'\vpp''}{3v_0^2} \,,
\end{equation}
where the pressure $P$ has been neglected because $P\ll \ve$ in the Newtonian limit. 
This can then be inserted into the scalar equation $J^r=0$, which yields
\begin{equation}\label{eq:WFphi}
\vpp'\left[\frac{2(5\s^2-2)\vpp'^2}{3v_0^2}-\frac{4GM}{r}-GM'\right]=0 \,.
\end{equation}
This equation has three branches of solutions: one with $\vpp'=0$, which gives identical predictions to 
GR, and two characterised by
\begin{equation}\label{eq:WFphisol2}
\frac{\vpp'^2}{v_0^2}=\frac{3}{5\s^2-2}\left(\frac{2GM}{r}+\frac{GM'}{2}\right),
\end{equation}
provided that $\s^2>2/5$. 

To select the correct branch with $\varphi \to 0$ far from the star, 
one needs to take into account the cosmological corrections in eqs. (\ref{eq:nusubh})-(\ref{eq:phisubh}) in order to determine the asymptotic form of the solutions.
Remarkably, as shown in the next section, one can find an exact solution of eqs. \eqref{E00}, \eqref{Err} and \eqref{eq:je0} outside the star. This will enable us to show that the correct solution is \cref{eq:WFphisol2} with $\vpp'<0$, by matching the exact solution with the sub-horizon weak-field approximation. It is not necessarily the case that an exact solution can be found for more general models and so we provide an alternate method for selecting the correct branch in appendix \ref{app:cosmoterms} for the reader wishing to study such models.

For now, we continue in the weak-field limit and substitute \cref{eq:WFphisol2} into the $00$- and $rr$- equations to find
\begin{align}
\dn'(r)&=\frac{6GM(r)}{(5\s^2-2)r^2}+\frac{(\s^2-1)GM''(r)}{2(5\s^2-2)}\,, \quad \textrm{and}\label{eq:dl1}\\
\dl(r)&=\frac{6GM(r)}{(5\s^2-2)r}-\frac{5(\s^2-1)GM'(r)}{2(5\s^2-2)r}.\label{eq:dl2}
\end{align}

Our next task is to calculate Newton's constant $\GN$. 
Since $\delta\nu/2 $ coincides with the potential $\Phi$ in the sub-horizon limit, it is immediate to identify $\GN$ by inspection of the first term on the right hand side of (\ref{eq:dl1}). One can also relate $\dl$ to the gravitational potential $\Psi$ defined in isotropic coordinates (see \cref{eq:iso}). 
In \cref{app:iso} we provide the coordinate transformation between the two coordinate systems and obtain the relation between $\dl$ and $\Psi$. 
Eventually, the equations \eqref{eq:dl1} and \eqref{eq:dl2} are found to be equivalent to 
\begin{align}
 \frac{\dd\Phi}{\dd r}&= \frac{\GN M}{r^2}+\frac{\Upsilon_1\GN M''}{4}\label{eq:dphiwf}\\
 \frac{\dd\Psi}{\dd r}& = \frac{\GN M}{r^2}-\frac{5\Upsilon_2\GN M'}{4r^2},\label{eq:dpsiwf}
 \end{align}
with
\begin{align}
\GN&=\frac{3G}{5\s^2-2}\label{eq:GN}\\
\Upsilon_1&=\Upsilon_2\equiv\Upsilon=-\frac{1}{3}\left(1-\s^2\right).\label{eq:ups}
\end{align}
Note that since $\s^2>2/5$ we have $-1/5<\Upsilon<\infty$. In this work we will focus our discussion mainly on the case $\U<0$, because it is less constrained  than the region $\U>0$  \cite{Sakstein:2015zoa,Sakstein:2015aac},  but we will show results for both regions for completeness. Outside the object one has $M''=M'=0$ and so one can see GR is recovered with the Eddington light bending (PPN) parameter $\gamma_{\rm PPN}=1$, confirming that the Newtonian limit of this theory agrees with GR outside extended sources. Equations \eqref{eq:dphiwf} and \eqref{eq:dpsiwf} constitute the main results of this subsection, supplemented by \eqref{eq:GN} and \eqref{eq:ups}. 

\subsection{Exact Vacuum Solution and Cosmological Matching}\label{ss:exact}

We now turn our attention to the full relativistic solution. Outside the stellar radius $R$ one has $\ve=P=0$, in which case the equations eqs. \eqref{E00}, \eqref{Err} and \eqref{eq:je0} admit the exact solution
\begin{align}
\nu(r)&=-\lambda(r)=\ln\left(1-\frac{\C}{r}-H^2r^2\right)\quad\textrm{and}\\
\phi(r)&=\pdot \left[t-\int \dd r \left(1-\frac{\C}{r}-H^2r^2\right)^{-1}\sqrt{\frac{\C}{r}+H^2r^2}\right],\label{eq:phiexact}
\end{align}
where the integration constant $\C$ must be found by matching to the interior solution at the stellar radius. 
This solution corresponds to exact Schwarzschild-de-Sitter metric with a non-trivial scalar field.
The presence of the nontrivial scalar field configuration results in a modification of the value of the cosmological constant.
Similar solutions were found previously in the case of the ``John'' term~\cite{Babichev:2013cya} and the quartic Horndeski term~\cite{Kobayashi:2014eva}.

Taking the sub-horizon limit $Hr\ll 1$ and assuming $\C/r\ll1$, i.e. the weak-field limit, one can match $\nu$ and $\lambda$ to find
\begin{equation}
\C=2 \GN M,
\end{equation}
independently of the branch of solutions chosen for $\vpp'$. Taking the weak-field limit in \cref{eq:phiexact} one finds
\begin{equation}
\frac{\phi'}{\pdot}=-\sqrt{\frac{2\GN M}{r}}
\end{equation}
showing that Vainshtein breaking branch with $\vpp'<0$ is the physical one. Furthermore, 
taking the limit $r\gg2\GN M$ and $r\sim H^{-1}$, we find that the metric potentials asymptote to their de Sitter forms \eqref{asymp_dS} so that this solution is fully consistent on all scales. We note that in the sub-horizon limit the metric is simply the Schwarzschild one, and so the PPN parameter $\beta_{\rm PPN}$ is unity. This means that the theory agrees with GR at the post-Newtonian level\footnote{Technically, it only agrees for effects such as the perihelion shift of Mercury. The other PPN parameters are not captured by the vacuum solution and require a more detailed modelling of the source (see \cite{Ip:2015qsa}). }.

\subsection{Compact Objects}

Having confirmed that the Vainshtein breaking branch is the physical one we now proceed to derive the TOV system of equations governing relativistic stars. Since we do not require Hubble-scale corrections we will 
concentrate on the quantities $\delta\nu$, $\delta\lambda$, and $\varphi$ introduced 
in equations \eqref{eq:nusubh}--\eqref{eq:phisubh} whereby we consider the corrections to the 
metric potentials and scalar due to the source and neglect terms of $\oo(Hr)$ and higher. In what follows, we work in units where $\GN=c=1$, which aids with the numerical integration.

We begin with the scalar equation of motion $J^r=0$, which gives
\begin{equation}
e^{-2\dl-\dn}r\left(\dl'+5\dn'\right)-4e^{-3\dl}\left(1+r\dn'\right)\frac{\vpp'^2}{\pdot^2}=0.
\end{equation}
This can be used to eliminate the scalar field perturbation from the $00$- and $rr$-equations. The final form of the equations is very long and not particularly enlightening and so we give them in \cref{app:tov} for completeness rather than presenting them here, although we note that after several manipulations they are of the first-order form $\dn'=g_1(\dn,\dl',\dl,\ve,\ve',P,P')$, $\dl'=g_2(\dn,\dl,\ve,\ve',P,P')$. The dependence on $\ve'$ indicates Vainshtein breaking.

In order to study the stability of stellar configurations it is useful to expand the functions near the centre as
\begin{align}
\dn(r)&=\dn_2\, r^2\\
\dl(r)&=\dl_2\, r^2\\
P(r)&=P_c+P_2\, r^2\\
\ve(r)&=\ve_c+\ve_2\, r^2,
\end{align}
where one can set $\dn(0)=\dl(0)=0$ by making a suitable coordinate redefinition and a subscript $c$ refers to the central density. Substituting these into the TOV system ($00$-, $rr$-, and equation \eqref{eq:emcons}) one finds
\begin{equation}
P_2=-\frac{\pi\GN}{3}(P_c+\ve_c)\left[3(2+5\U)P_c+(2+3\U)\ve_c\right].
\end{equation}
Physically acceptable stellar solutions can only be obtained if $P''(0)<0$ \cite{Delgaty:1998uy}, which imposes the condition
\begin{equation}
\U>-\frac{2}{3}\left(1+3\frac{P_c}{\ve_c}\right)\left(1+5\frac{P_c}{\ve_c}\right)^{-1}.
\end{equation}
In the non-relativistic limit $P_c/\ve_c\ll1$ this reduces to the lower limit found by \cite{Saito:2015fza} (using the same method). The extra terms in $P_c/\ve_c$ represent relativistic corrections to this. In particular, in the ultra-relativistic limit $P_c=\ve_c$ one has $\U>-4/9\approx -0.44$. We note that our model restricts $\U $ to be larger than this and so the unstable region of parameter space is not accessible. Furthermore, this bound may change in more general models (for example if one were to include a cubic galileon).

\subsubsection{Neutron Stars}

\begin{figure}[ht]\centering
\includegraphics[width=0.49\textwidth]{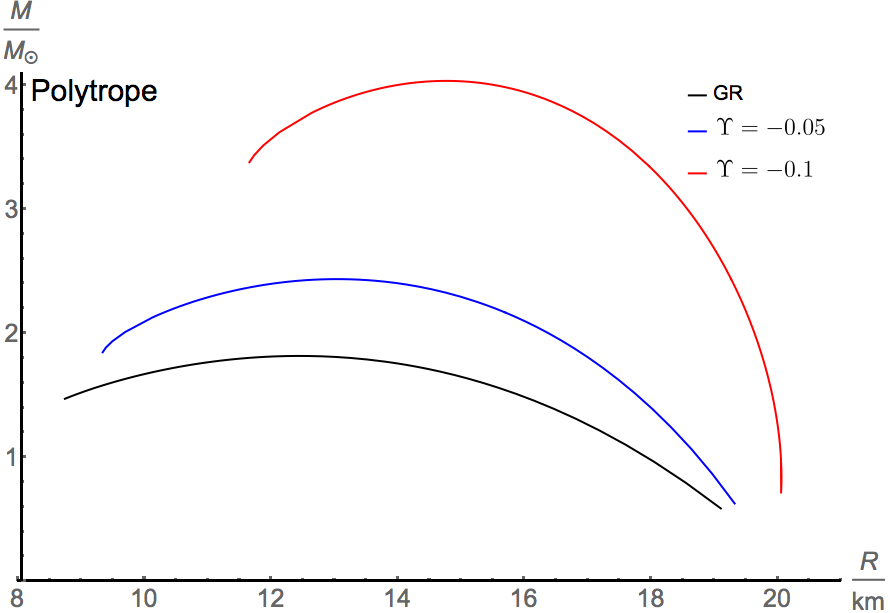}
\includegraphics[width=0.49\textwidth]{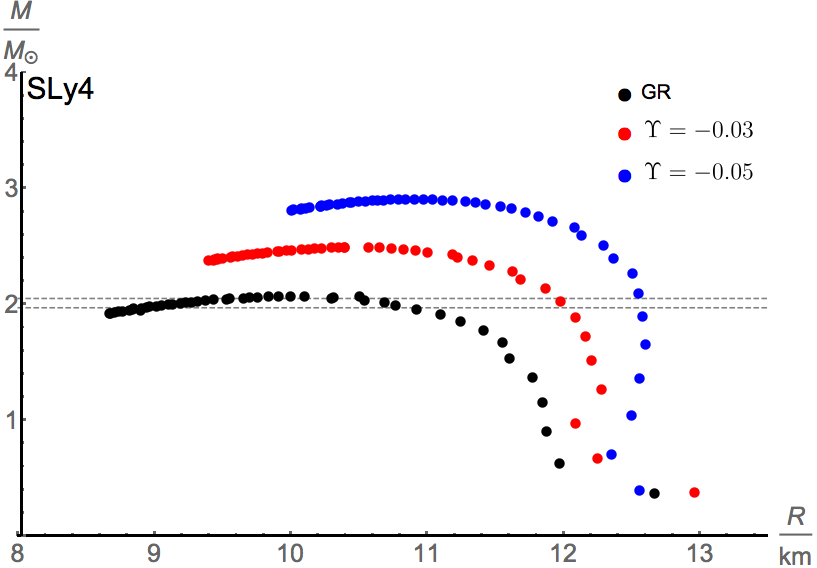}
\vspace{2cm}
\includegraphics[width=0.49\textwidth]{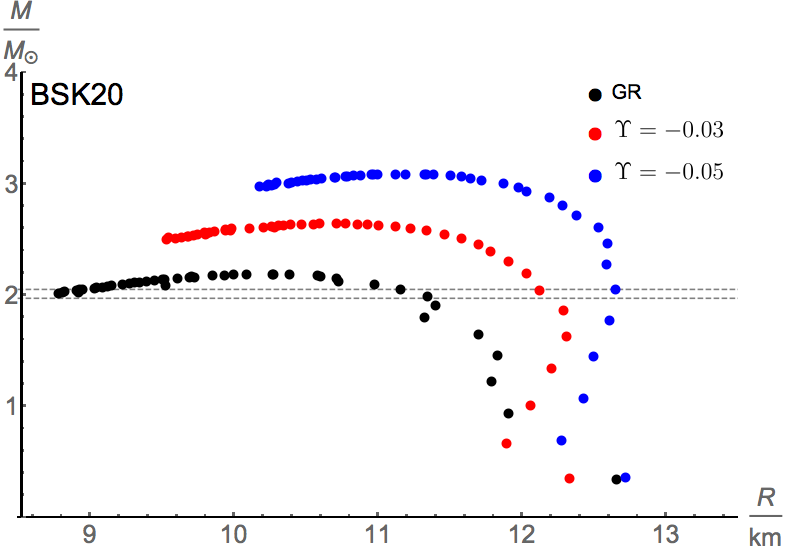}
\includegraphics[width=0.49\textwidth]{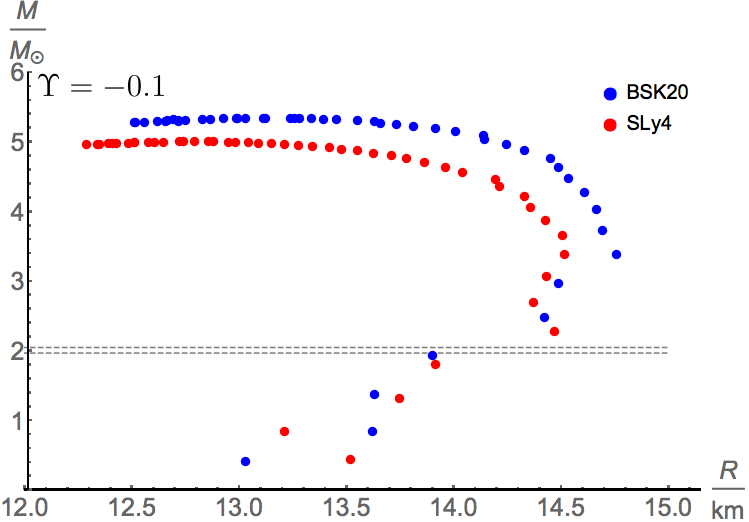}
\caption{The mass-radius relation for the polytropic model (top left), SLy4 EOS (top right panel), BSK20 EOS (bottom left panel) for varying values of $\U<0$ indicated in the plots, and the extreme case $\U=-0.1$ using the SLy4 and BSK20 equations of state (bottom right panel). The region between the gray dashed lines represents the highest mass neutron star observed ($M=2.01\pm0.04M_\odot$). Note that the axes on each plot have different scales.}\label{fig:real}
\end{figure}

The TOV system consisting of the $00$-, $rr$- and energy-momentum conservation equations do not close and one needs to provide an equation of state. We will first consider the polytropic equation of state 
\begin{equation}\label{eq:poly}
\ve=\left(\frac{P}{K}\right)^{\frac{1}{2}}+P,
\end{equation}
with $K=123M_\odot^2$ which corresponds to a fluid with $P=K\rho^2$ where $\rho$ is the matter density. This equation of state is not particularly realistic although it gives compact objects and is useful to facilitate a comparison with other works looking at compact objects in alternative gravity theories such as \cite{Cisterna:2015yla,Cisterna:2016vdx,Silva:2016smx,Maselli:2016gxk}. For this reason, we show the mass-radius relation for this equation of state in the top left panel of \cref{fig:real} without any observational data as a tool to see how beyond Horndeski theories compare with GR. Curves for GR and beyond Horndeski theories with $\U=-0.05$ and $\U=-0.1$ are shown. One can see that more extreme modifications of GR (more negative $\U$) result in larger radii at fixed mass and a larger maximum mass. We will see that this is a generic feature that is retained when more realistic equations of state are used. 

Indeed, in order to compare with observations one requires a more realistic EOS coming from nuclear theory and currently there is a wide range of candidate equations of state that are consistent with observations to varying extents \cite{Lattimer:2012nd}. These typically come in tabulated form or can be calculated directly using coupling coefficients coming from various nuclear physics calculations. In our theory, we have the added complication that one must also specify the derivative of the pressure ($\dd P/\dd\ve$) and so it is convenient to use an EOS for which analytic fits are available. For this reason, we use two realistic equations of state: SLy4 \cite{Douchin:2001sv} (see \cite{Haensel:2004nu} for the fitting function) and BSK20 \cite{Goriely:2010bm,Pearson:2012hz} (see \cite{Potekhin:2013qqa} for the fitting function).

In \cref{fig:real} we plot the mass-radius relation for both the SLY4 and BSK20 equations of state when the theory of gravity is GR and our beyond Horndeski model
 with $\U=-0.03$ and $\U=-0.05$. One can see qualitatively similar features to the polytropic case, namely a higher maximum mass and a shift to larger radii. Observationally, the most massive neutron star thus far observed is PSR J0348+0432 with a mass $M=2.01\pm0.04M_\odot$ \cite{Demorest:2010bx}, and both 
equations of state give stable neutron stars that are consistent with this observation when the theory of gravity is GR. One can see that even mild deviations from GR ($\U=-0.05$) predict stars as massive as $3M_\odot$. Of course, such predictions are consistent with the highest mass observed neutron star and so one may hope to get constraints by looking at smaller mass objects. Indeed, one can see that 
our model with $\U<0$ predicts radii that can be 1 km or larger than the GR prediction at fixed mass. Typically, fits to neutron star masses and radii predict radii less than 14 km at 2$\s$ for masses between $1$ and $2M_\odot$ \cite{Heinke:2014xaa} and one can see that the neutron stars predicted by the parameter range studied are consistent with this. Moving to larger values of $\U$, we plot the (not so) extreme value $\U=-0.1$ in the bottom right panel of \cref{fig:real} where drastic deviations from GR can be seen; the masses can be as large as 5 $M_\odot$ (or larger) in stark contrast with the current stellar evolution paradigm where the cores of massive stars collapse to form black holes. For lower mass objects, radii larger than 14 km are predicted, which are in slight tension with reported observations \cite{Heinke:2014xaa} although it is difficult to rule these values out with certainty because the allowed region in the $M$--$R$ plane depends on the assumed atmosphere model\footnote{The study in \cite{Heinke:2014xaa} considers both hydrogen and helium atmospheres although they note that carbon atmospheres would allow for even larger radii.} and the distance to the globular cluster where the objects are observed, which may change in modified gravity models \cite{Jain:2012tn,Vikram:2014uza}.

\begin{figure}[ht]\centering
\includegraphics[width=0.49\textwidth]{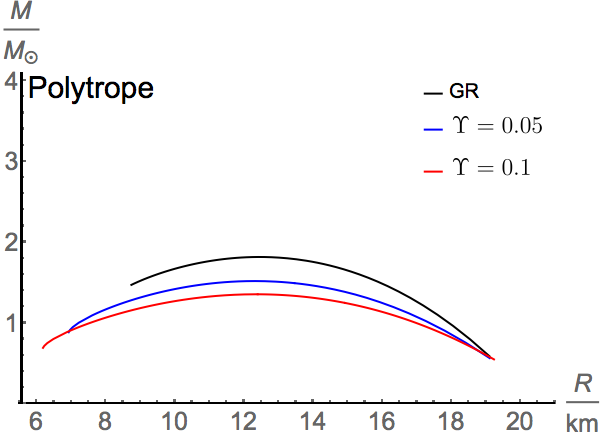}
\includegraphics[width=0.49\textwidth]{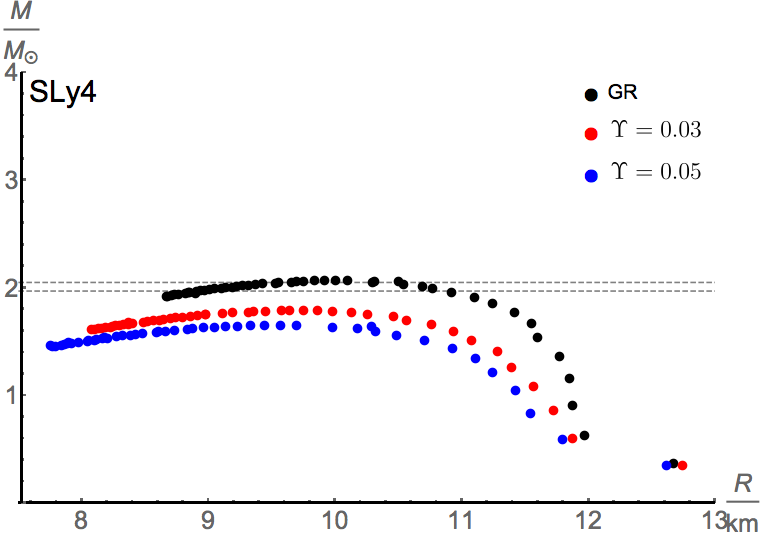}
\vspace{2cm}
\includegraphics[width=0.49\textwidth]{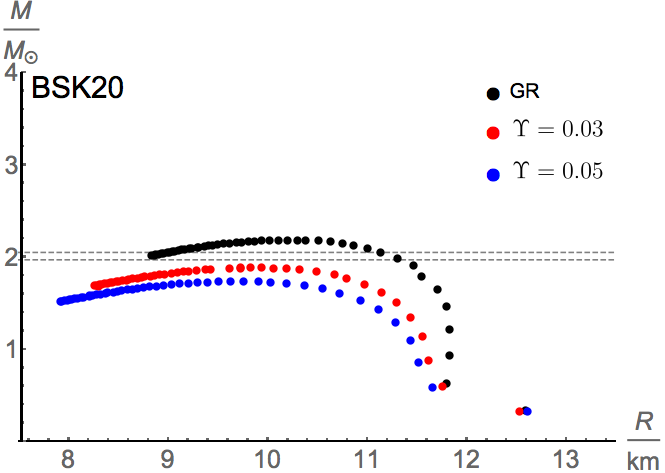}
\includegraphics[width=0.49\textwidth]{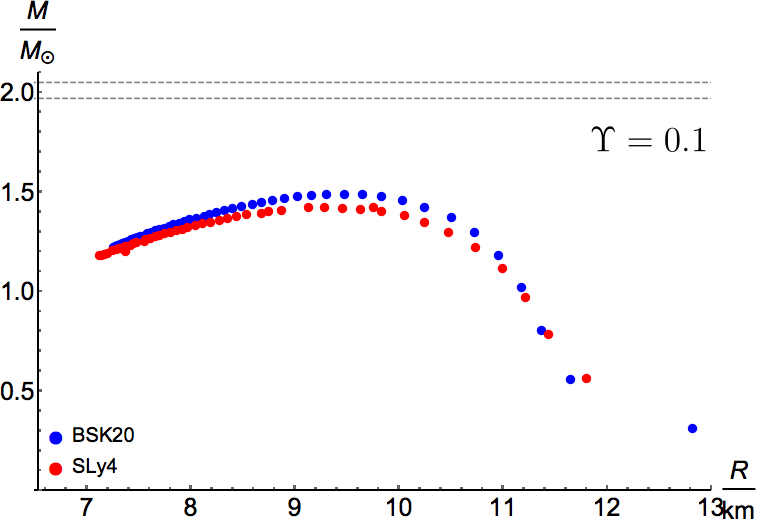}
\caption{The mass-radius relation for the polytropic model (top left), SLy4 EOS (top right panel), BSK20 EOS (bottom left panel) for varying values of $\U>0$ indicated in the plots, and the extreme case $\U=-0.1$ using the SLy4 and BSK20 equations of state (bottom right panel). The region between the gray dashed lines represents the highest mass neutron star observed ($M=2.01\pm0.04M_\odot$). Note that the axes on each plot have different scales}\label{fig:posU}
\end{figure}

In figure \ref{fig:posU} we plot the mass-radius relations for the same equations of state using positive values of $\U$ indicated in the caption. As one would expect, the behaviour is opposite to $\U<0$ i.e. larger $\U$ predicts lower maximum masses and smaller radii at fixed mass. Small deviations from GR in this direction may be in tension with PSR J0348+0432 since the predicted maximum mass is less then $2 M_\odot$. However, whether or not this can be used to rule out such values is unclear since other equations of state that produce masses far in excess of $2M_\odot$ when the theory is GR, such as MPA1, MS1 and AP3 \cite{Ozel:2016oaf}, may yield masses compatible with PSR J0348+0432 for the same values of $\U$.

\subsubsection{White Dwarf Stars}

\begin{figure}[h]\centering
\includegraphics[width=0.7\textwidth]{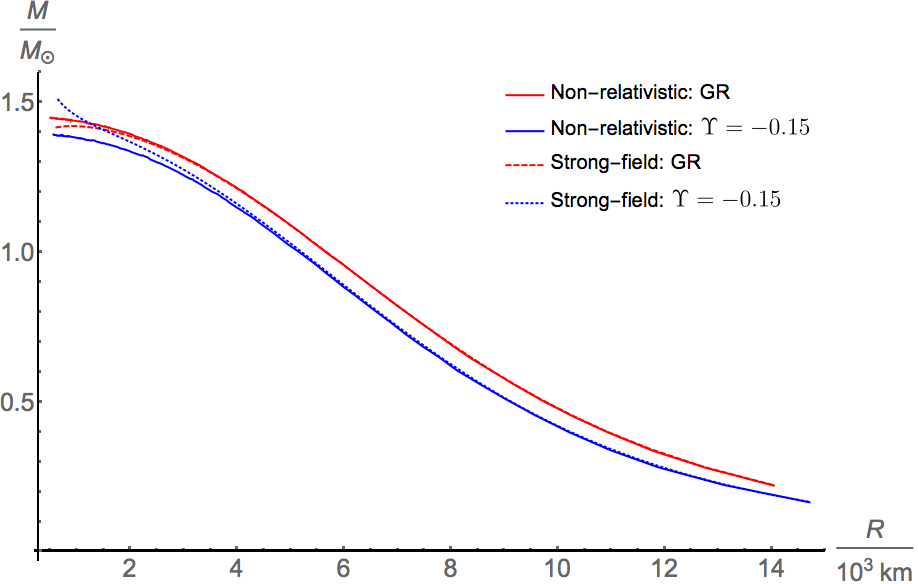}
\caption{The mass-radius relation for white dwarf stars. The solid curves show the predictions found using the weak-field limit and the dashed curves show the predictions found by solving the full TOV system. GR predictions are shown using red curves and beyond Horndeski with $\U=-0.15$ are shown in blue.}\label{fig:WD}
\end{figure}

Before concluding, we revisit white dwarf stars in these theories, which were previously studied by \cite{Jain:2015edg} using the non-relativistic limit. They were able to obtain the bounds $-0.22<\U<0.54$, the lower limit coming from the Chandrasekhar mass and the upper from fitting the mass-radius relation. While it is well-known in GR that strong-field effects can be ignored for white dwarf stars (one has $GM/R\sim 10^{-6}$ and so the weak-field limit is appropriate) this remains to be seen for beyond Horndeski theories, especially for the case $\U<0$ which may result in more compact objects so that post-Newtonian effects are important. We will use the same equation of state as \cite{Jain:2015edg}, who consider the white dwarf as a non-interacting gas of completely ionized $^{12}$C (see \cite{Jain:2015edg} for the details). In \cref{fig:WD} we compare the mass-radius relation for white dwarfs found using the weak- and strong-field equations for both GR and for $\U=-0.15$. One can see that the GR curves are very close with small deviations for very compact objects. In contrast, the beyond Horndeski relations differ greatly for compact configurations and crosses the GR prediction, predicting more massive objects than GR. 

Since \cite{Jain:2015edg} use lower mass objects to obtain bounds using the mass-radius relation we expect these to be robust but it is important to check that the different shapes at the high-mass end do not affect the $\chi^2$ fit. Conversely, the lower bound of $-0.22$ was obtained by comparing the predicted Chandrasekhar mass with the highest mass white dwarf presently observed. Since strong field effects raise this prediction above the GR value we conclude that this bound is not robust and so the bound is weakened to the one coming from the mass-radius relation, namely $-0.48\le\U\le0.54$\footnote{One can impose $\-0.18\le\U\le0.27$ at 1$\sigma$.}. In fact, our lower bound $\U\gsim-0.44$ supersedes this.

\section{Discussion and Conclusions}\label{sec:concs}

In this paper we have studied relativistic objects in beyond Horndeski theories of gravity focusing on the simplest model that exhibits Vainshtein breaking. This allowed us to derive the non-relativistic limit of this theory in a fully self-consistent and relativistic manner that is free from coordinate ambiguities and includes Hubble-scale effects. We were able to find an exact analytic solution for the object's exterior space-time that is Schwarzschild-de Sitter, which allowed us to confirm that the Vainshtein breaking branch is the physical one. Furthermore, this shows that the PPN parameter $\beta_{\rm PPN}=1$ so that the theory agrees with GR at the post-Newtonian level.

Turning our attention to the structure of relativistic stars, we derived the Tolman-Oppenheimer-Volkoff equations and used them to find a relativistic correction to the lower bound on $\uo$ found by \cite{Saito:2015fza} that must be satisfied in order to have static spherically symmetric stellar configurations. In particular, the bound is raised from $\uo>-2/3$ to $\uo>-4/9$. We solved the TOV equations for polytropic and two realistic neutron star equations of state (SLy4 and BSK20) and consistently found 
mass-radii relations with larger maximum masses and larger radii at fixed mass than predicted by GR when $\uo<0$. Configurations with $\uo\gsim-0.05$ (note that $\uo$ closer to zero implies smaller deviations from GR) predict maximum masses $M\sim3M_\odot$ and radii $R\lsim $14 km favoured by observation \cite{Heinke:2014xaa} whereas $\uo\lsim-0.05$ predict masses that can be larger than $5M_\odot$ and radii in excess of 14 km.

We are unable to claim any specific constraints since masses in excess of $2M_\odot$ are technically consistent with observations of the highest mass neutron star\footnote{ By this we mean that predictions for the maximum mass higher than $2M_\odot$ do not immediately rule out the model because the theory also predicts $2M_\odot$ stars and it may be the case that heavier objects exist and have not been observed. Future observations of the radius of PSR J0348+0432 may help to constrain the parameter $\U$ because we generically find larger radii at fixed mass but this may not be so straightforward due to degeneracies with the equation of state.} \cite{Demorest:2010bx} and uncertainties with atmospheric models and distances to globular clusters (which can be affected by modified gravity) may allow for larger radii. 

Another important issue worth mentioning is that of discriminating between different modified gravity theories. Many theories exhibit the Vainshtein mechanism (without the partial breaking) and therefore give identical predictions to GR but other theories such as the Fab Four \cite{Charmousis:2011bf,Cisterna:2015yla,Maselli:2016gxk} and theories that exhibit spontaneous scalarization \cite{Damour:1993hw,Minamitsuji:2016hkk} also predict deviations in the mass-radius relation. One then has two simultaneous problems: degeneracy with the equation of state and other modified gravity models. The first can be addressed by looking for equation of state-independent relations, such as the relation between the dimensionless moment of inertia ($\bar{I}=Ic^2/\GN^2M^3$) and the compactness $\GN M/Rc^2$), which \cite{Breu:2016ufb} have recently found to be approximately universal in GR. For example,  \cite{Maselli:2016gxk} discuss this relation in the context of the Fab Four theory and find a universal but different relation for a certain subset (John) but the same relation for a different subset (Ringo). A measurement of this relation could therefore distinguish between different modified gravity theories without the complications coming from the equation of state. The equations we have used here are considerably longer and their derivation considerably more technical than the other theories mentioned here and so clearly a calculation of the $\bar{I}$--$\mathcal{C}$ relation is beyond the scope of the present paper; we intend to investigate this in future work.

Despite the lack of a firm constraint, our investigation has been fruitful: it is now clear that the Vainshtein breaking branch can be physical and that the theory satisfies post-Newtonian tests of gravity. This makes beyond Horndeski theories viable alternatives to $\Lambda$CDM that make novel predictions inside astrophysical bodies whilst still satisfying traditional tests of GR. Furthermore, we have derived a new lower bound on the range of allowed parameters and have been able to show that neutron star solutions can be found with masses and radii that are consistent with observations. 

Finally, we re-examined the structure of white dwarf stars and showed that post-Newtonian corrections are important for high mass stars and therefore the bounds coming from white dwarf tests may need to be revisited.

\section*{Acknowledgements}

We are grateful to Wynn Ho and Masato Minamitsuji for several enlightening discussions. EB was supported in part by the research program, Programme national de cosmologie et galaxies of the CNRS/INSU, France and Russian Foundation for Basic Research Grant No. RFBR 15-02-05038. RS was supported in part by JSPS postdoctoral fellowships for research abroad.

\appendix

\section{Relation Between Schwarzschild and Newtonian Metric Potentials}\label{app:iso}

In this appendix we show the Schwarzschild potentials $\dl$ and $\dn$ are related to the Newtonian potentials $\Phi$ and $\Psi$ in the weak field limit and neglecting cosmological corrections (i.e. $(Hr)^2\ll\delta\nu,\,\delta\lambda$). The two coordinate systems are
\begin{align}
\dd s^2 &= -(1+\dn(r))\dd t^2 + (1+\dl(r))\dd r^2 + r^2\dd\Omega_2^2\quad\textrm{Schwarzschild}\\
\dd s^2 &= (-1+2\Phi(\rt))\dd t^2 +(1+2\Psi(\rt))\left(\dd\rt^2+\rt^2\dd\Omega_2^2\right)\quad\textrm{Isotropic},
\end{align}
where we use the same coordinate $t$ in both systems since only the radial coordinate varies. For this reason, one can instantly read off the relation $\Phi(\rt)=-\delta\nu(r)/2$, where we will see below that $\rt=r$ to Newtonian order. The transformation 
\begin{equation}
 r=\left(1+\Psi(\tilde{r})\right)\tilde{r}
\end{equation}
with 
\begin{equation}
 \frac{\dd\Psi}{\dd \tilde{r}}=-\frac{\delta\lambda(\tilde{r})}{\tilde{r}}
\end{equation}
brings the Schwarzschild metric into isotropic form.

\section{Perturbative Cosmological Matching}\label{app:cosmoterms}

As noted above, the exact Schwarzchild-de Sitter solution we have found for our specific model may not be general for all beyond Horndeski theories, in which case one needs to find the correct branch by matching to the physical solution at asymptotic infinity. In this appendix, we give an example of the method for doing this by feigning ignorance of the exact solution and performing the perturbative matching explicitly. One may straightforwardly apply this procedure to more complicated theories.

Previously, we ignored cosmological effects by taking the limit $H^2r^2 \ll \delta \nu, \delta \lambda$ and $v_0H r^2/2 \ll \varphi$, as well as the weak-field limit $GM/r\ll1$. Here, we still consider the weak-field limit and the sub-horizon limit ($Hr \ll 1$) but we now retain the leading-order terms in $Hr$\footnote{Note that there are two sources of cosmological corrections to the equations. The first come from \cref{eq:nusubh}--\cref{eq:phisubh} and arise because the asymptotic space-time is de Sitter. The second come from $k_2$ and $f_4$ because they are related to $H$ via the cosmological (Friedmann) equations of motion (see \cref{eq:f4k2rel}). } so that the tensor and scalar equations of motion lead to\footnote{Note that we are working at distances larger than the stellar radius so that $M\equiv M(R)$ and $M'=M''=0$.}
	\begin{align}\label{eq:WFphiH}
		\left( \frac{\varphi'}{v_0} - Hr \right)\left\{ \left( \frac{\varphi'}{v_0} - Hr \right)^2 - \frac{2G_N M}{r} - (Hr)^2 \right\} = 0 \,,
	\end{align}
with solutions
\begin{align}
	\frac{\varphi'}{v_0} &= Hr \quad\textrm{and}\\ \frac{\varphi'}{v_0}& = Hr \pm \sqrt{\frac{2G_N M}{r} + (Hr)^2}\approx Hr(1\pm1)\pm\frac{\GN M}{Hr^2} \,,
\end{align}
where the approximation applies for $r\gg(\GN M/H^2)^{1/3}$. One can see that \cref{eq:WFphiH} reproduces \cref{eq:WFphi} for $v_0H r^2/2 \ll \varphi$ and $H^2 r^2 \ll 2G_N M/r \sim \delta \nu\,, \delta \lambda$. Only the solution with $\vpp'<0$ decays at large distances whereas the other two grow as $Hr$, confirming that this is indeed the branch that matches onto the asymptotic cosmological space-time and is therefore the physical one. Furthermore, linearising \cref{eq:WFphiH} (i.e. ignoring terms cubic and quadratic in $\vpp'$) one finds that the linear solution at large distances is $\vpp'/\pdot = -\GN M/H r^2$. The branch with $\vpp'<0$ approaches this asymptotically---which one expects in galileon theories---and this is typically how the correct branch is chosen in models defined in the decoupling limit.

The method we have presented here generalises in a straight-forward manner to more complicated theories. One expects longer equations and extra parameters (because only two parameters in the Lagrangian can be replaced by $H$ and $\pdot$) but, since $\varphi'=0$ is a solution at all orders in $Hr$ when $M=0$, one can always choose the correct branch by expanding the sourced equations with respect to $Hr$ until the degeneracy is resolved.
\section{TOV System}\label{app:tov}

Here we present the equations for $\dl'$ and $\dn'$ for completeness. The equation for $\dn'$ is relatively simple and is of the form
\begin{equation}
\dn'=\frac{4 e^{\delta \lambda }-8 \pi G_{\rm N}  e^{\delta \lambda } P r^2 (5 \Upsilon +1) (\dl' r-5)+8 \pi G_{\rm N} e^{\delta \lambda } r^2 (5 \Upsilon +1) \varepsilon -e^{\delta \lambda } \dl' r-4}{r \left(e^{\delta \lambda }-8 \pi G_{\rm N} e^{\delta \lambda } r^2 (5 \Upsilon +1) \varepsilon +4\right)}\end{equation}
whereas the equation for $\dl'$ is of the form
\begin{equation}
\dl'=-\frac{U}{V}
\end{equation}
with
\begin{align}
U&=-e^{2 (\delta \lambda +\delta \nu )}+e^{3 \delta \lambda +2 \delta \nu }+5 e^{\delta \lambda } \Upsilon +32 \pi ^3 e^{2 \delta \lambda } P^3 r^6 G_{\rm N}{}^3 (5 \Upsilon +1)^3 \left(16 e^{\delta \lambda +2 \delta \nu }-16 e^{2 \delta \nu }-25 \Upsilon \right)\nonumber\\&+12 \pi ^2 e^{\delta \lambda } P^2 r^4 G_{\rm N}{}^2 (5 \Upsilon +1)^2 \left(16 e^{2 (\delta \lambda +\delta \nu )}-16 e^{\delta \lambda +2 \delta \nu }-15 e^{\delta \lambda } \Upsilon +40 \Upsilon \right)\nonumber\\&-2 \pi e^{\delta \lambda } r^2 G_{\rm N} (5 \Upsilon +1) \varepsilon \left(4 e^{2 (\delta \lambda +\delta \nu )}+3 e^{\delta \lambda } \Upsilon +2048 \pi ^3 P^3 r^6 G_{\rm N}{}^3 (5 \Upsilon +1)^3 e^{2 (\delta \lambda +\delta \nu )}\right.\nonumber\\&\left.+96 \pi ^2 e^{\delta \lambda } P^2 r^4 G_{\rm N}{}^2 (5 \Upsilon +1)^2 \left(8 e^{\delta \lambda +2 \delta \nu }+5 \Upsilon \right)\right.\nonumber\\&\left.+4 \pi P r^2 G_{\rm N} (5 \Upsilon +1) \left(24 e^{2 (\delta \lambda +\delta \nu )}+29 e^{\delta \lambda } \Upsilon +16 \pi e^{\delta \lambda } r^3 G_{\rm N} \Upsilon (5 \Upsilon +1) \varepsilon'-4 \Upsilon \right)\right.\nonumber\\&\left.+80 \pi e^{\delta \lambda } P' r^3 G_{\rm N} \Upsilon ^2+16 \pi e^{\delta \lambda } P' r^3 G_{\rm N} \Upsilon +320 \pi P' r^3 G_{\rm N} \Upsilon ^2+64 \pi P' r^3 G_{\rm N} \Upsilon +40 \pi e^{\delta \lambda } r^3 G_{\rm N} \Upsilon ^2 \varepsilon'\right.\nonumber\\&\left.+8 \pi e^{\delta \lambda } r^3 G_{\rm N} \Upsilon \varepsilon'-18 \Upsilon \right)\nonumber\\&+4 \pi ^2 e^{2 \delta \lambda } r^4 G_{\rm N}{}^2 \Upsilon (5 \Upsilon +1)^2 \varepsilon ^2 \left(-40 \pi P r^2 G_{\rm N} (5 \Upsilon +1)+32 \pi P' r^3 G_{\rm N} (5 \Upsilon +1)-13\right)\nonumber\\&+2 \pi P r^2 G_{\rm N} (5 \Upsilon +1) \left(3 \left(-4 e^{2 (\delta \lambda +\delta \nu )}+4 e^{3 \delta \lambda +2 \delta \nu }+22 e^{\delta \lambda } \Upsilon -e^{2 \delta \lambda } \Upsilon +4 \Upsilon \right)\right.\nonumber\\&\left.+8 \pi e^{\delta \lambda } \left(e^{\delta \lambda }+4\right) r^3 G_{\rm N} \Upsilon (5 \Upsilon +1) \varepsilon'\right)+80 \pi e^{\delta \lambda } P' r^3 G_{\rm N} \Upsilon ^2+10 \pi e^{2 \delta \lambda } P' r^3 G_{\rm N} \Upsilon ^2\nonumber\\&+16 \pi e^{\delta \lambda } P' r^3 G_{\rm N} \Upsilon +2 \pi e^{2 \delta \lambda } P' r^3 G_{\rm N} \Upsilon +160 \pi P' r^3 G_{\rm N} \Upsilon ^2+32 \pi P' r^3 G_{\rm N} \Upsilon +40 \pi e^{\delta \lambda } r^3 G_{\rm N} \Upsilon ^2 \varepsilon'\nonumber\\&+10 \pi e^{2 \delta \lambda } r^3 G_{\rm N} \Upsilon ^2 \varepsilon'+8 \pi e^{\delta \lambda } r^3 G_{\rm N} \Upsilon \varepsilon'+2 \pi e^{2 \delta \lambda } r^3 G_{\rm N} \Upsilon \varepsilon'-5 \Upsilon\end{align}
and
\begin{align}
V&=r \left(8 \pi P r^2 G_{\rm N} (5 \Upsilon +1)+1\right) \left(e^{2 (\delta \lambda +\delta \nu )}+4 \pi ^2 e^{2 \delta \lambda } P^2 r^4 G_{\rm N}{}^2 (5 \Upsilon +1)^2 \left(16 e^{2 \delta \nu }+5 \Upsilon \right)
\right.\nonumber\\&\left.+2 \pi e^{\delta \lambda } P r^2 G_{\rm N} (5 \Upsilon +1) \left(8 e^{\delta \lambda +2 \delta \nu }+e^{\delta \lambda } \Upsilon -6 \Upsilon \right)\right.\nonumber\\&\left.+2 \pi e^{\delta \lambda } r^2 G_{\rm N} \Upsilon (5 \Upsilon +1) \varepsilon \left(e^{\delta \lambda }+12 \pi e^{\delta \lambda } P r^2 G_{\rm N} (5 \Upsilon +1)-6\right)+4 \pi ^2 e^{2 \delta \lambda } r^4 G_{\rm N}{}^2 \Upsilon (5 \Upsilon +1)^2 \varepsilon ^2+5 \Upsilon \right).
\end{align}

\bibliographystyle{JHEP}
\bibliography{ref_2}

\end{document}